\documentclass[amsmath,pra,twocolumn]{revtex4}
\usepackage{times}
\usepackage{epsfig}
\usepackage{amssymb}
\usepackage{amsmath}
\usepackage{amsfonts}
\usepackage{bm}

\begin{document}

\title{Importance of nonresonant corrections for the description of atomic spectra}
\author{D. Solovyev$^1$, A. Anikin$^{1}$, T. Zalialiutdinov$^{1}$ and L. Labzowsky$^{1,2}$ }

\affiliation{ 
$^1$ Department of Physics, St. Petersburg
State University, Petrodvorets, Oulianovskaya 1, 198504,
St. Petersburg, Russia
\\
$^2$ Petersburg Nuclear Physics Institute, 188300, Gatchina, St. Petersburg, Russia}

\begin{abstract}
We demonstrate that the present superaccurate measurements of transition processes between atomic states in hydrogen atom reached the limit of accuracy when transition frequency cannot be defined anymore in a unique way. This was predicted earlier and is due to the necessity to include the nonresonant corrections in the description of resonant processes. The observed spectral line profile becomes asymmetric, and it becomes impossible to extract the value of transition frequency from this profile in a unique way. Nonresonant corrections depend on the type of experiment and on the experimental arrangement. However, the line profile itself for any resonant process can be defined with any desired level of accuracy. A popular trend in modern search for atomic frequency standards and atomic clocks is the search for transitions where the nonresonant corrections are negligible. In this paper we present closed expressions for the resonant photon scattering cross sections on an atomic level with dependence on all atomic quantum numbers including fine and hyperfine structure. These expressions are given for different types of experiments with fixing of the incident (outgoing) photon propagation directions and incident (outgoing) photon polarization. Using these expressions we demonstrate that the transition frequencies in particular cases cannot be derived uniquely if the accuracy of measurement reaches the level quoted in  A. Beyer {\it et al}. Our interpretation of the results of this experiment is alternative to the interpretation given by A. Beyer {\it et al}.
\end{abstract}

\maketitle
\section{Introduction}

This paper is devoted to the interpretation of the results of the recent experiment \cite{1}. During the last decades the accuracy of spectroscopic measurements for hydrogen atom has grown up considerebly and now reaches 15 digits for the $1s-2s$ transition frequency \cite{2}. A question arises whether the improvement of accuracy of resonant transition frequency measurement can be endless or it is limited by some reasons. Here we ignore such problems as Doppler, collisional and blackbody broadening. One can imagine an experiment with a single cold isolated atom, when the spectral line profile will be purely natural. This question was answered in \cite{3,4,5} on the basis of quantum qlectrodynamics (QED). The term "nonresonant corrections" (NR corrections) was introduced in \cite{3,4}. In \cite{5} the same words (NR corrections) were employed and the reference to \cite{4} was given.

Later in \cite{5a,5b,5c,5d} on the basis of QED and quantum mechanics (QM) NR corrections were treated under the name of "quantum interference effects" (QIE). The QIE include a very broad set of quantum effects, for example quantum beats. The NR corrections represent a particular type of QIE, namely a distortion of the spectral line shape in the resonance spectroscopy of atoms and molecules. This type of QIE was first discussed  on the basis of QED in \cite{3,4,5}.

A QED theory of atomic spectral line profile was first developed by F. Low \cite{6} wherefrom the existence of the nonresonant corrections also followed. Unlike the resonant value of transitions frequency, the NR corrections depend on the process of excitation of atomic level, on the type of experiment and on the method of extracting of transition frequency value from the experimental data. Therefore the refinement of the transition frequency value may have sense until the NR corrections are smaller than the accuracy of experiment \cite{3,4}. For all cases investigated in \cite{3,4,5} and later works on the subject the NR corrections appeared to be negligible. In particular, according to \cite{7} this was the case also for the two-photon transition frequency measurement for the $1s-2s$ transition in hydrogen in \cite{2}. The situation changed when the results of the highly accurate measurement of the transition frequencies $2s_{1/2}^{F=0}\rightarrow 4p_{1/2}^{F=1}$ and $2s_{1/2}^{F=0}\rightarrow 4p_{3/2}^{F=1}$ were reported in \cite{1}. The uncertainty of these measurement was quoted to be considerably smaller than the observed interference effects. According to the line profile theory these interference effects manifest the existence of NR corrections. 
In the present paper we investigate the problem from this point of view. We derive expressions for the cross section of resonant photon scattering on hydrogen atom with the fine and hyperfine structure taken into account. These expressions contain dependence on the directions and polarizations of the incident (absorbed) and outgoing (emitted) photons. This allows one to describe different types of experiments with different correlations between directions and polarizations of both photons. All these results are then applied to derivation of NR corrections to the photon scattering cross sections and transition frequencies. 

We focus on NR corrections originating from the neighboring fine structure level components as in \cite{1} where mutual influence of transitions $2s_{1/2}^{F=0}\rightarrow 4p_{1/2}^{F=1}$ and $ 2s_{1/2}^{F=0}\rightarrow 4p_{3/2}^{F=1} $ was observed. First we consider NR corrections to $2s_{1/2}^{F=0}\rightarrow 4p_{1/2}^{F=1}$ transitions due to the quantum interference with $2s_{1/2}^{F=0}\rightarrow 4p_{3/2}^{F=1}$ transitions. Corresponding corrections to another transition $2s_{1/2}^{F=0}\rightarrow 4p_{3/2}^{F=1}$ are similar but have an opposite sign. We demonstrate that the NR corrections to these transitions in this particular case do not depend on the type of experiment and on the experimental geometry. However, they depend on the choice of the detected decay branch: they are different when the detection process ends up in the states with $ F=0,\,1$ or $2 $. When the frequency of the outgoing photon is not fixed at all, the result of the measurement begins to depend both on the type of the experiment and on the experimental arrangement (geometry).

In the present paper we imagine an experimental set-up for measuring the transition frequencies in an atom (very simplified and schematic) as follows. The atoms located in some volume are excited by the laser beam with changeable frequency. The direction of propagation of this beam also can be varied. The laser photons (incident photons) excite the atomic levels via the process of inelastic photon scattering. The line shape corresponding to this process is the source for determining the atomic transition frequencies. This information is gained by observing the decay of excited levels (outgoing photons). The outgoing photons are registered by some detector; the position of this detector defines the direction of the outgoing photons. In our paper we consider experiments of two types. In an experiment of the first type the directions of propagation for both the incident photon (coinciding with direction of the laser beam) and the outgoing photon (defined by the position of detector) are fixed. Then the cross section depends on the angle between these directions of propagations of both photons. In an experiment of the second type the incident photon is polarized, and its direction is arbitrary. In this case cross section depends on the angle between the incident photon polarization and the outgoing photon direction.

The experiment \cite{1} was of the last type and the decay branch was not fixed. Combining the results of different measurements, the authors of \cite{1} found the way to cancel NR corrections. The frequency corresponding to this combination they interpreted as a $2s_{1/2}^{F=0}\rightarrow 4p_{1/2}^{F=1}$ frequency. In the same way the $2s_{1/2}^{F=0}\rightarrow 4p_{3/2}^{F=1}$  transition frequency was determined.

This interpretation was later supported by the more detailed analysis in \cite{5d}. In \cite{5d} it was also mentioned that the interpretation given in \cite{1} is equivalent to the use of "magic angles" (see the definition of "magic angles" below in section IV). In the present paper we discuss another interpretation of the result of the experiment \cite{1}. Even with the use of "magic angles" that annihilate the dependence of transition frequencies on experimental geometry, the dependence on the choice of the final decay channels remains. This dependence is given explicitly in Eq. (\ref{three}) below. The values of transition frequency in Eq. (\ref{three}) include NR corrections which in this particular case do not depend on geometry but depend on the final decay channel. This supports the statement made earlier in \cite{3,4} that the transition frequency between atomic levels cannot be defined uniquely with any desired accuracy.

\section{QED theory of resonant photon scattering on atomic electron with account for the fine and hyperfine level structure}
\label{section2}

For an accurate description of NR corrections to the atomic transition frequencies it is natural to employ the QED theory of atomic processes developed in particular in \cite{8,9}. The resonant scattering corresponds to the case when the incident photon frequency is chosen to be equal to the difference of atomic level energies $ \omega=E_n-E_i $ for the particular $ n $ value. We denote by $ i$, $n$, $f$ the initial, intermediate and final atomic states, respectively. Then in the sum over intermediate states in the scattering amplitude only one term for the chosen $ n $ value should be retained. According to the QED theory of spectral line profile \cite{6,8,9} in case of resonance an infinite set of Feynman graphs containing electron self-energy insertions in the electron line $ n $
should be taken into account. This leads to the arrival of the level width $ \Gamma_{n} $ in the energy denominator corresponding to the resonant state $ n $ of the scattering amplitude. Taking into account the fine and hyperfine structure of atomic levels we will further understand the indices $ i,\,n,\,f $ as standard sets of quantum numbers: principal quantum number $ n $, electron orbital angular momentum $ l $, electron total electron momentum $ j $, atomic angular momentum $ F $, and its projection $ M_{F} $. 

With these notations, the photon scattering amplitude in the nonrelativistic limit and in the resonant approximation is \cite{10}
\begin{gather}
\label{3}
U=(E_{nljF}-E_{n_i l_i j_i F_i})^{3/2}(E_{nljF}-E_{n_f l_f j_f F_f})^{3/2}
\\\nonumber
\times
\sum_{M_F}\frac{ \langle n_i l_i j_i F_i M_{F_i} | \vec{e}_{i}\vec{d}  |n l j F M_F \rangle  }{E_{nljF}-E_{n_il_ij_iF_i}-\omega-\frac{\mathrm{i}\Gamma}{2}}
\\\nonumber\times
\langle n l j F M_F | \vec{e}_{f}^{\;*}\vec{d}  |n_f l_f j_f F_f M_{F_f} \rangle  
,
\end{gather}
where $ \Gamma=\Gamma_{nljF} $. In Eq. (\ref{3}), $  \vec{e}_{i} $, $  \vec{e}_{f} $ are the polarization vectors of the incident and emitted photons, $ \vec{d}=e\vec{r} $ is the operator of the electric dipole moment of the electron, $ e $ is the electron charge. Relativistic units are employed: $ \hbar=c=1 $. 

The scattering amplitude apart from a term given by Eq. (\ref{3}) contains another term with interchanged absorption and emission photons. This term does not contribute to the resonant scattering. Below we will take into account the dominant nonresonant contribution arising from Eq. (\ref{3}).

The cross section of the resonant photon scattering reads 
\begin{eqnarray}
\label{4}
\sigma_{if}=\frac{(E_{nljF}-E_{n_i l_i j_i F_i})^3(E_{nljF}-E_{n_f l_f j_f F_f})^3}{2F_{i}+1}
\\\nonumber
\times\sum_{M_{F_i}M_{F_f}}
\left|\sum_{M_F}\frac{\langle n_i l_i j_i F_i M_{F_i} | \vec{e}_{i}\vec{d}  |n l j F M_F \rangle   }{(E_{nljF}-E_{n_il_ij_iF_i}-\omega)^2+\frac{\Gamma^2}{4}}\right.
\\\nonumber\times
\left.  \langle n l j F M_F | \vec{e}_{f}^{\;*}\vec{d}  |n_f l_f j_f F_f M_{F_f} \rangle
\right|^2
d\omega
.
\end{eqnarray}
Here we have summed over atomic angular momentum projections in the final state and averaged over the atomic angular momentum projections of initial state. In Eqs. (\ref{3}) and (\ref{4}) we restricted ourselves with the most important case of emission (absorption) of E1 photons. 

\section{Application to the description of different types of experiments on the measurement of atomic transition frequencies}

Eq. (\ref{4}) for the cross section of the resonant photon scattering on an atomic electron is general and can be used to describe any experiment involving this process. We will concentrate on the experiment for observation of the spectral line shape of a transition $ n_i l_i j_i F_i \rightarrow nljF$ and extracting the transition frequency from the experimental data. This corresponds to the experiment \cite{1}. We will distinguish two types of experiments of that sort. In an experiment of the first type the directions of photon propagation are fixed: the incident photon direction $ \vec{\nu}_{i} $ coincides with the direction of the laser beam and the outgoing photon direction $ \vec{\nu}_f $ is defined by the detector position. In the second type of experiment the incident photon polarization $ \vec{e}_i $ and the outgoing photon direction $ \vec{\nu}_f $ are fixed; this is exactly the situation in experiment \cite{1}. In the nonrelativistic limit the matrix elements in Eq. (\ref{4}) do not depend explicitly on the photon directions $ \vec{\nu}_{i} $ and $ \vec{\nu}_{f} $. Implicitly this dependence enters via the transversality condition. Dependence on $ \vec{\nu}_{i} $, $ \vec{\nu}_{f} $ becomes explicit after summation over photon polarizations. Then for the type 1 experiment we have to evaluate $ \sum\limits_{\vec{e}_i, \vec{e}_f}\sigma_{if} $, for the the type 2 experiment it is necessary to evaluate $ \sum\limits_{\vec{e}_f}\sigma_{if} $. 

To introduce the NR correction to the cross section given by Eq. (\ref{4}) we have to take into account the next term of the scattering amplitude, closest by energy to the resonant term. The set of a quantum numbers for this additional state should allow connection of this state with the initial state by absorption of photon in electric dipole transitions. Therefore the neighbouring fine structure components of the resonant level may give a noticeable NR correction as it was recently observed in \cite{1}. In what follows we will consider the NR corrections originating from the states with the same $ nl $ quantum numbers as the resonant state, but different values of $ j$ and $F $. We will neglect the contribution quadratic in NR correction and will neglect the level width in the energy denominator corresponding to the NR state. 

We represent the cross section in the form
$\sigma_{if}=\sigma^{\mathrm{res}}_{if}+\sigma^{\mathrm{nr}}_{if}$, where $ \sigma^{\mathrm{res}}_{if} $ denotes now the resonant contribution and $ \sigma^{\mathrm{nr}}_{if} $ represents the NR correction. In the NR correction we retain only the interference term between the resonant and nonresonant amplitudes. For evaluating the cross section with Eq. (\ref{4}) we employ the techniques of irreducible tensor operators (we follow notations given in \cite{11}). After lengthy but standard evaluations and after summation over all angular momenta projections we arrive at the following expressions (see Appendix for the derivation). For the experiment of the type 1,
\begin{gather}
\label{13}
\sum_{\vec{e}_i, \vec{e}_f}\sigma_{if}^{\rm nr} =2\mathrm{Re}
\sum_{\substack{nljF \\ n'l'j'F'}}
(E_{nljF}-E_{n_i l_i j_i F_i})^{3/2}
\\\nonumber
\times
(E_{n'l'j'F'}-E_{n_i l_i j_i F_i})^{3/2}
(E_{nljF}-E_{n_f l_f j_f F_f})^{3/2}
\\\nonumber
\times(E_{n'l'j'F'}-E_{n_f l_f j_f F_f})^{3/2}\sum_{xy}
A_{xy}^{(1)}
\\\nonumber
\times
\frac{
\left\lbrace
\left\lbrace \nu^{i}_1\otimes\nu^{f}_1  \right\rbrace_{y}\otimes \left\lbrace \nu^{i}_1\otimes\nu^{f}_1  \right\rbrace_{y}
\right\rbrace_{00}
d\omega }{(E_{nljF}-E_{n_il_ij_iF_i}-\omega-\frac{\mathrm{i\Gamma}}{2})(E_{nlj'F'}-E_{n_il_ij_iF_i}-\omega)}
.
\end{gather}
Here $ \nu^{i}_1  $, $\nu^{f}_1 $ denote the irreducible tensors of the rank 1 corresponding to the vectors $ \vec{\nu}_i $, $ \vec{\nu}_f $ respectively in the laboratory frame, symbol $ \otimes $ denotes a tensor product and 
\begin{gather}
\label{14}
A_{xy}^{(1)}=
\frac{36(-1)^{F'-F+x-y}}{2F_{i}+1}
\Pi_{x}^2\Pi_{y}
\begin{Bmatrix}
1 & 1 & y \\
1 & 1 & x
\end{Bmatrix}
\begin{Bmatrix}
1 & x & 1 \\
1 & 1 & 1
\end{Bmatrix}
^2
\\\nonumber\times
\begin{Bmatrix}
1 & x & 1 \\
F' & F_{i} & F
\end{Bmatrix}
\begin{Bmatrix}
1 & x & 1 \\
F' & F_{f} & F
\end{Bmatrix}
\\\nonumber
\times
\langle n_i l_i j_i F_i ||d_{1}||n l j F \rangle 
\langle n' l' j' F' ||d_{1}||n_i l_i j_i F_i\rangle 
\\\nonumber
\times
\langle n_f l_f j_f F_f||d_{1}||n' l' j' F'\rangle 
\langle n l j F ||d_{1}||n_f l_f j_f F_f \rangle 
\end{gather}
where $ \Pi_{a}=\sqrt{2a+1} $. The reduced matrix element of the dipole operator in Eq. (\ref{14}) is given by \cite{11}
\begin{gather}
\label{red}
\langle n'l'j'F'||d_1||nljF\rangle= (-1)^{j'+j+I+l'+1/2+F}
\\\nonumber
\times
\Pi_{F'}\Pi_{F}\Pi_{j'}\Pi_{j}
\begin{Bmatrix}
j' & F' & I \\
F  & j  & 1
\end{Bmatrix}
\begin{Bmatrix}
l' & j' & 1/2 \\
j  & l  & 1
\end{Bmatrix}
\langle n' l' || d_{1} || nl \rangle,
\end{gather}
where $ I $ is the nuclear spin ($ I=1/2 $ for hydrogen atom) and
\begin{eqnarray}
\label{rad}
\langle n' l' || d_{1} || nl \rangle = e(-1)^{l'}\Pi_{l}\Pi_{l'}
\begin{pmatrix}
l & 1 & l'\\
0 & 0 & 0
\end{pmatrix}
\\\nonumber
\times
\int_{0}^{\infty}r^3 R_{n'l'}R_{nl}dr.
\end{eqnarray}
Here $ R_{nl} $ is the radial part of hydrogen wave function. Similar evaluations for an experiment of the type 2 yield
\begin{gather}
\label{15}
\sum_{\vec{e}_f}\sigma_{if}^{\rm nr} =
2\mathrm{Re}
\sum_{\substack{nljF \\ n'l'j'F'}}
(E_{nljF}-E_{n_i l_i j_i F_i})^{3/2}
\\\nonumber
\times
(E_{n'l'j'F'}-E_{n_i l_i j_i F_i})^{3/2}
(E_{nljF}-E_{n_f l_f j_f F_f})^{3/2}
\\\nonumber
\times
(E_{n'l'j'F'}-E_{n_f l_f j_f F_f})^{3/2}
\sum_{xy}A_{xy}^{(2)}
\\\nonumber
\times
\frac{
\left\lbrace
\left\lbrace e^{i}_1\otimes\nu^{f}_1  \right\rbrace_{y}\otimes \left\lbrace e^{i}_1\otimes\nu^{f}_1  \right\rbrace_{y}
\right\rbrace_{00}
d\omega}{(E_{nljF}-E_{n_il_ij_iF_i}-\omega-\frac{\mathrm{i}}{2}\Gamma)(E_{nlj'F'}-E_{n_il_ij_iF_i}-\omega)}
 , 
\end{gather}
where
\begin{gather}
\label{16}
A_{xy}^{(2)}=\frac{6(-1)^{F'-F-y}}{2F_{i}+1}\Pi_{x}^2\Pi_{y}
\\\nonumber\times
\begin{Bmatrix}
1 & 1 & y \\
1 & 1 & x
\end{Bmatrix}
\begin{Bmatrix}
1 & 1 & x \\
1 & 1 & 1
\end{Bmatrix}
\begin{Bmatrix}
1 & x & 1 \\
F' & F_{i} & F
\end{Bmatrix}
\begin{Bmatrix}
1 & x & 1 \\
F' & F_{f} & F
\end{Bmatrix}
\\\nonumber
\times
\langle n_i l_i j_i F_i ||d_{1}||n l j F \rangle 
\langle n' l' j' F' ||d_{1}||n_i l_i j_i F_i\rangle 
\\\nonumber
\times
\langle n_f l_f j_f F_f||d_{1}||n' l' j' F'\rangle 
\langle n l j F ||d_{1}||n_f l_f j_f F_f \rangle 
.
\end{gather}
Tensor product in Eq. (\ref{13}) can be expressed through trigonometric functions of the angle between the vectors $ \vec{\nu}_i $, $ \vec{\nu}_f $ (see Appendix). In Eq. (\ref{15}) $ e^{i}_{1} $ is the irreducible tensor of the rank $ 1 $ corresponding to the vector $ \vec{e}_i $. The summation over $ x $ in Eq. (\ref{13}) and the summation over $ y $ in Eq. (\ref{15}) run over the values $ x=0,\,1,\,2 $ and $y=0,\,1,\,2$, respectively. The resonant contributions $\sum\limits_{\vec{e}_i, \vec{e}_f}\sigma_{if}^{\rm res}$ and $\sum\limits_{\vec{e}_f}\sigma_{if}^{\rm res}$ to the total cross section are derived from Eqs. (\ref{13})-(\ref{16}) by setting $ n'l'j'F'=nljF $ with insertion of level widths in the resonant denominator. In Appendix we use the notations $ A^{(1,2)\;\mathrm{res} }_{xy} $ and $ A^{(1,2)\;\mathrm{nr}}_{xy} $ for the resonant and nonresonant contributions to the photon scattering cross section respectively.

\section{Determination of transition frequency}

The dependence of the photon scattering cross section $ \sum\limits_{\vec{e}_i,\vec{e}_f}\sigma_{if} $, $ \sum\limits_{\vec{e}_f}\sigma_{if} $ on the incident photon frequency represents the natural line profile for the transition $ n_i l_i j_i F_i \rightarrow n l j F$. In this paper we neglect all other types of line broadening. Qualitatively, our conclusions will remain valid for any type of line profile (Voigt and Gauss) though the numerical values for transition frequencies may slightly change. 

The resonant transition frequency $ \omega_{\rm res} $ can be defined from $ \sigma_{if}(\omega) $ by different ways. One evident way is to define $ \omega_{\rm res} $ as $\omega_{\rm res}=\omega_{\rm max}$, where $ \omega_{\rm max} $ corresponds to the maximum value of $ \sigma_{if}(\omega) $. Then $ \omega_{\rm res} $ can be obtained from the condition
\begin{eqnarray}
\label{18}
\frac{d}{d\omega}\sigma_{if}(\omega)=0.
\end{eqnarray}
In the resonant approximation we immediately find
\begin{eqnarray}
\label{19}
\omega_{\rm res}=\omega_{\rm max}=\omega_{0}=E_{nljF}-E_{n_il_i j_i F_i}.
\end{eqnarray}
As long as the line profile remains symmetric with respect to $ \omega=\omega_{\rm max} $, the definition (\ref{19}) remains the same for any other way of extracting $ \omega_{\rm res} $ from the line profile. Experimentally, the line center is commonly defined by the fitting procedure. In principle, the line center may not coincide with the maximum of the line contour, so that Eq. (\ref{18}) is not always applicable. However, no other way to define theoretically the transition frequency is yet known and it is natural to use for this purpose the simple definition given by Eq. (\ref{18}), which gives the correct answer in the absence of NR corrections \cite{3,4,5}.

For both types of experiments discussed above the expression for the photon scattering cross section can be parametrized in the form:
\begin{widetext}
\begin{eqnarray}
\label{main}
\sigma_{if}^{(1,2)}=C\left[\frac{f^{(1,2)}_{\rm res}}{(\omega_0-\omega)^2+\frac{\Gamma^2}{4}}
+2\mathrm{Re}\frac{f^{(1,2)}_{\rm nr}}{(\omega_0-\omega-\frac{\mathrm{i}\Gamma}{2})\Delta} \right]
d\omega
,
\end{eqnarray}
\end{widetext}
where $ \Delta=E_{nlj'F'}-E_{nljF} $,
$\sigma^{(1)}_{if}=\sum\limits_{\vec{e}_i,\vec{e}_f}\sigma_{if}$, $\sigma^{(2)}_{if}=\sum\limits_{\vec{e}_f}\sigma_{if}$, $ C $ is some constant that is not important for our further derivations and
\begin{widetext}
\begin{eqnarray}
\label{final}
f^{(1,2)}_{\rm res}=\left[(E_{nljF}-E_{n_il_ij_iF_i})(E_{nljF}-E_{n_fl_fj_fF_f}) \right]^{3}
\sum_{xy}A_{xy}^{(1,2)\;\mathrm{res}}\left\lbrace\left\lbrace a_{1}^{(1,2)}\otimes b_{1}^{(1,2)} \right\rbrace_{y} 
\otimes
\left\lbrace a_{1}^{(1,2)}\otimes b_{1}^{(1,2)} \right\rbrace_{y}\right\rbrace_{00},
\end{eqnarray}
\begin{eqnarray}
\label{final2}
f^{(1,2)}_{\rm nr}=\left[(E_{nljF}-E_{n_il_ij_iF_i})(E_{nlj'F'}-E_{n_il_ij_iF_i})
(E_{nljF}-E_{n_fl_fj_fF_f})(E_{nlj'F'}-E_{n_fl_fj_fF_f}) \right]^{3/2}
\\\nonumber
\times
\sum_{xy}A_{xy}^{(1,2)\;\mathrm{nr}}\left\lbrace\left\lbrace a_{1}^{(1,2)}\otimes b_{1}^{(1,2)} \right\rbrace_{y} 
\otimes
\left\lbrace a_{1}^{(1,2)}\otimes b_{1}^{(1,2)} \right\rbrace_{y}\right\rbrace_{00}.
\end{eqnarray}
\end{widetext}
Coefficients $ A^{(1,2)}_{xy} $ are defined by Eqs. (\ref{14}) and (\ref{16}), with $ a^{(1)}_1=\nu_{1}^{i} $, $ a^{(2)}_1=e_{1}^{i} $, $b^{(1)}_1=b^{(2)}_1=\nu_{1}^{f}$. Using the definition of $ \omega_{\rm res} $ via the "maximum" of the line profile according to Eq. (\ref{18}) we find 
\begin{widetext}
\begin{eqnarray}
\label{add1}
\frac{d}{d\omega}\sigma_{if}(\omega)
=
-\frac{8
\left(
f^{(1,2)}_{\mathrm{nr}}
\left(
\Gamma^2-4 (\omega-\omega_{0})^2
\right)+4 \Delta  f^{(1,2)}_{\mathrm{res}} (\omega-\omega_{0})
\right)
}
{\Delta  \left(\Gamma^2+4 (\omega-\omega_{0})^2\right)^2}
=0
.
\end{eqnarray}
\end{widetext}
Expansion of Eq. (\ref{add1}) into the Taylor series in the vicinity of $ \omega_{0} $
yields
\begin{eqnarray}
\label{add2}
-\frac{8 f^{(1,2)}_{\mathrm{nr}}}{\Gamma^2\Delta}
-\frac{32 f^{(1,2)}_{\mathrm{res}}(\omega -\omega_{0})}{\Gamma^4}
+O\left((\omega -\omega_{0})^2\right)
=0
.
\end{eqnarray}
Finally, neglecting the terms of the order $ O\left((\omega -\omega_{0})^2\right) $ in Eq. (\ref{add2}) and solving it with respect to $ \omega $ we arrive at the definition of $ \omega_{\rm max} $ 
\begin{eqnarray}
\label{23}
\omega_{\rm max}^{(1,2)}=\omega_{0}-\delta\omega^{(1,2)},
\end{eqnarray}
where
\begin{eqnarray}
\label{24}
\delta\omega^{(1,2)} = \frac{f^{(1,2)}_{\rm nr}}{f^{(1,2)}_{\rm res}}\frac{\Gamma^2}{4\Delta}.
\end{eqnarray}
With our definition of $ \Delta $ this value corresponds to the lower component of the fine structure of the level $ nl $. For the upper sublevel of two neighboring components of energy level we would arrive to the same expression as Eq. (\ref{23}) but with thw opposite sign of $ \Delta $ and with $ \Gamma=\Gamma_{nlj'F'} $. NR correction in Eq. (\ref{23}) can depend on the arrangement of the experiment ,i.e., on the angles between the vectors $ \nu_{i} $ and $ \nu_{f} $ in the experiment of type 1 or on the angles between the vectors $ \vec{e}_i $ and $ \vec{\nu}_f $ in the experiment of type 2. 

The smallness of NR corrections in Eq. (\ref{24}) is defined by the ratio $ \Gamma /\Delta $. Eq. (\ref{24}) is obtained as the lowest term of an expansion of the result in terms of $ \Gamma /\Delta $. The approximations that were used for derivation of Eq. (\ref{4}) are valid up to the higher order terms in parameter $ \Gamma /\Delta $. This parameter is always small for two neighbouring components of the fine structure (see below the particular example below). The parameter $ \Gamma /\Delta $ may not be small for two neighbouring hyperfine sublevels, but this requires special investigation \cite{7}.

\section{Application to $ 2s_{1/2}^{F=0}\rightarrow 4p_{1/2}^{F=1} $ and $ 2s_{1/2}^{F=0}\rightarrow 4p_{3/2}^{F=1} $ transitions}

Now we turn to evaluation of $ 2s_{1/2}^{F=0}\rightarrow 4p_{1/2}^{F=1} $ transition frequency with account for NR corrections originating from the neighboring $ 4p_{3/2}^{F=1} $ level. For this purpose we set in all equations $ n_i l_i=2s $, $ j_i=1/2 $, $ F_i=0 $, $ nl=4p $, $ j=1/2 $, $ F=1 $, $ j'=3/2 $, $ F'=1 $. As the final states we have chosen states listed in Table \ref{tab1}. Note that hyperfine structure of $ 1s $ and $ 2s $ electron shells was resolvable in experiments \cite{1}. The results of evaluations are presented in Table \ref{tab1}. 
\begin{table}
\caption{The NR corrections in kHz to the transitions frequency $ 2s_{1/2}^{F=0}\rightarrow 4p_{1/2}^{F=1} $ with the account for the neighbouring $ 4p_{3/2}^{F=1} $ state for the experiment of the type 2 ($ \vec{e}_i\vec{\nu}_f $ correlation). The same values are obtained for the experiment of the type 1 ($ \vec{\nu}_i\vec{\nu}_f $ correlation).}
\begin{tabular}{|c c|}
\hline
Final state & $ \delta \omega^{(2)} $\\
\hline
$ 1s_{1/2}^{F=0} $ &   61.2355\\
$ 1s_{1/2}^{F=1} $ &  -30.6178 \\
$ 2s_{1/2}^{F=0} $ &   61.2357 \\
$ 2s_{1/2}^{F=1} $ &  -30.6178 \\
$ 3s_{1/2}^{F=0} $ &   61.2362 \\
$ 3s_{1/2}^{F=1} $ &  -30.6181 \\
$ 3d_{3/2}^{F=1} $ &   30.6174\\
$ 3d_{3/2}^{F=2} $ &    6.1236  \\
\hline
\end{tabular}
\label{tab1}
\end{table}
For evaluation of NR corrections according to Eqs. (\ref{main}), (\ref{final}), (\ref{24}) we use theoretical values given in \cite{5c}, which incorporate relativistic, QED, nuclear size, the hyperfine structure corrections. The same concerns the value of the widths $ \Gamma=\Gamma_{4p_{1/2}^{F=1}}=1.2941\times 10^{7} $ Hz and the fine structure interval $ \Delta = E_{4p_{3/2}^{F=1}}-E_{4p_{1/2}^{F=1}}= 1367433.3$ kHz \cite{5c}. These values give a sufficiently accurate result for $ \delta \omega $ up to four digits after the decimal point. The parameter $ \Gamma /\Delta $ in this case is equal to $ 0.00946 $, so the expansion in powers of this parameter works very well.

As can be seen from Table \ref{tab1}, the NR corrections to the transition frequency $ 2s_{1/2}^{F=0}\rightarrow 4p_{1/2}^{F=1} $ do not depend on the type of experiment and consequently on the geometry of this experiment. However, these NR corrections appear to depend strongly on the method of the frequency detection, i.e. on the choice of the state to which the excited $ 4p_{1/2}^{F=1} $ level finally decays. Moreover, this dependence concerns only the quantum numbers of this final state, and the result is nearly independent on the frequency of the outgoing photon. The latter circumstance is understandable since according to Eq. (\ref{24}) the NR corrections are proportional to the ratio $ f_{\rm nr}/f_{\rm res}$ where the corresponding energy differences nearly cancel. 

When the hyperfine structure of the final levels is resolved, the NR corrections differ only by the values of the total angular momentum $ F_{f} $ of the final hyperfine sublevel. This can be seen from the closed expressions (\ref{14}), (\ref{16}), (\ref{main}), (\ref{final}) for the NR corrections via $ 6j $-symbols. Therefore, for the transition frequency $ 2s_{1/2}^{F=0}\rightarrow 4p_{1/2}^{F=1} $, three different values of $ \omega_{\rm res}^{\rm max\;(1,2)} $ corresponding to $ F_f=0,\,1,\,2 $ can be derived for both types of experiment by using $ \omega_0 $ from \cite{5c} and NR corrections from Table \ref{tab1}:
\begin{eqnarray}
\label{three}
F_f=0\;\;\;\omega_{\rm res}^{\rm max\;(1,2)}
=616520152497.3\;\mathrm{kHz}
\\\nonumber
F_f=1\;\;\;\omega_{\rm res}^{\rm max\;(1,2)}
=616520152527.9\;\mathrm{kHz}
\\\nonumber
F_f=2\;\;\;\omega_{\rm res}^{\rm max\;(1,2)}
=616520152552.4\;\mathrm{kHz}
\end{eqnarray}

These three values differ from each other by more than 50 kHz. This is 15 times larger than the accuracy of measurement quoted in \cite{1} (3 kHz). Nevertheless, all 3 numbers in Eq. (\ref{three}) have equal rights to be interpreted as "$ 2s_{1/2}^{F=0}-4p_{1/2}^{F=1} $ transition frequency". If in the process of the frequency measurement only the emission of the outgoing photon is detected without fixing of its frequency, the summation over all the final states should be done. In the case of our interests this summation looks as follows
\begin{eqnarray}
\label{avr}
\delta\omega^{(1,2)} = \frac{\sum\limits_{n_fl_fj_fF_f}f^{(1,2)}_{\rm nr}}{\sum\limits_{n_fl_fj_fF_f}f^{(1,2)}_{\rm res}}\frac{\Gamma^2}{4\Delta}
.
\end{eqnarray}

\begin{figure}[hbtp]
\caption{The NR correction for the transition $ 2s_{1/2}^{F=0}\rightarrow 4p_{1/2}^{F=1} $ as the function of the angle between the vectors $ \vec{\nu}_i $, $ \vec{\nu}_f $ for the experiment of the type 1 (solid line) and as the function of the angle between the vectors $ \vec{e}_i $, $ \vec{\nu}_f $ in the experiment of the type 2 (dashed line) according to Eq. (\ref{avr}).}
\centering
\includegraphics[scale=0.85]{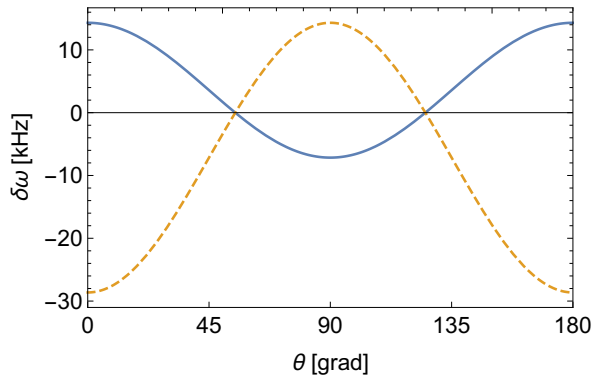}
\label{fig2}
\end{figure}

Now the NR correction begins to depend on the type of the experiment and on the angles between the vectors $ \vec{\nu}_i $, $ \vec{\nu}_f $ in the experiment of the first type or between the vectors $ \vec{e}_i $, $ \vec{\nu}_f $ in the experiment of the second type. The results for $ 2s_{1/2}^{F=0}\rightarrow 4p_{1/2}^{F=1}  $ transition are presented in Fig. \ref{fig2}.

According to Eq. (\ref{avr}) and Fig. \ref{fig2}, the NR correction vanishes for certain angles $ \theta_1=54.7^{\circ}  $ and $ \theta_2=125.3^{\circ}  $, which are the same for both types of experiment. The possible use of the "magic angles" for determination of transition frequencies in atoms was mentioned in \cite{5c,5d}. In \cite{5d} it was noted that the method of extracting the transition frequency value from the experimental data used in \cite{1} is actually equivalent to the use of "magic angles". The same "magic angles" arise in different areas of quantum physics where the interference of two electric dipole amplitudes is involved, see for example \cite{nmr}. The "magic angles" are connected with the roots of equation $ P_{2}(x) =0$ where $ P_{2} $ is the Legendre polynomial (see details in the Appendix).

Recently, evaluation of atomic transition frequencies with the use of "magic angles" was considered in \cite{amaro1}. Values for "magic angles" in \cite{amaro1} coincide with quoted above for similar transitions. Evaluation of transition frequency $ 2s_{1/2}^{F=0}-4p_{1/2}^{F=1} $ with the use of Eq. (\ref{avr}) for "magic angles" with the theoretical values $ \omega_0 $, $ \Gamma $ and $ \Delta $ from \cite{5c} gives  
\begin{eqnarray}
\label{new1}
\omega_{\rm res}^{\rm max\;(1,2)}=616520152558.5\;\mathrm{kHz}.
\end{eqnarray}
A similar evaluation of the $2s_{1/2}^{F=0}-4p_{3/2}^{F=1}  $ transition frequency yields
\begin{eqnarray}
\label{new2}
\omega_{\rm res}^{\rm max\;(1,2)}=616521519991.8\;\mathrm{kHz}.
\end{eqnarray}

As we understand the main goal of the work [1] is not to present the new frequency standard but to present an accurate experimental result which can be conveniently compared to the theory for extraction some fundamental constants. For this purpose, the determination of a certain characteristics of experimental line profile was employed. The determination of this characteristics which in [1] was called "frequency" included summation over the decay channels, using "magic angles" and then fitting to the theoretical line shape. An advantage of this recipe is that it fully avoids the QIE and that the "frequency" can be directly compared to the theoretical parameter $\omega_ {0} $” (difference of the energy levels for free atom). Another advantage is that this recipe looks to be universal: pursuing the same goal one can for any transition in any atom to sum over decay channels and to use "magic angles" geometry. Our goal in this paper is to demonstrate that the recipe used in [1] and hence the definition of frequency is not unique. We suggest another recipe: choose a certain decay channel then the dependence on the experimental geometry vanishes automatically. Then the maximum of the experimental line shape can be compared with the theoretical line shape where NR effects (QIE) should be included. In principle, this is more complicated but also quite realizable way for extracting fundamental constants from the comparison of experiment and theory. The frequencies defined in this way are given in Eq. (\ref{three}). These frequencies correspond to the energy differences not in a free atom, but in the atom, perturbed by the measurement. The universality of the recipe given in [1] also is not absolute; if the final state has many decay channels (e.g. as the Rydberg state) it may be more convenient to choose a certain decay channel and to employ the frequency values of the type Eq. (\ref{three}).”

\section{Conclusions}

An analysis performed in this paper demonstrates that the frequencies of certain transitions (in particular $2s_{1/2}^{F=0}\rightarrow 4p_{1/2}^{F=1}$ transition) in hydrogen atom cannot be defined uniquely at the high accuracy of the recent measurement. This happens due to the presence of NR corrections depending on the type of experiment, on the experimental arrangement and on the method of extracting the transition frequency values from the experimental data. Since the influence of the NR corrections is unavoidable, the same situation will occur for any atomic transition as soon as the accuracy of measurement will become comparable with NR corrections. 

The recent numerous works on the improvement of frequency standards and on the construction of new atomic clocks are based on the search for atomic transitions where NR corrections are negligible. However, it is important to remember that the limit for the accuracy of transition frequency measurement does exist for any atomic transition. 

The statement that the excited atomic states are not directly observable was made originally in \cite{3,4}. Strictly speaking, observable are only the line profiles of various processes, such as for the light emission and absorption by atoms, scattering of light and different particles on atoms. A theoretical picture of an atom with fine and hyperfine level structure is a very elaborate, accurate and useful model (but only a model) for description of all this variety of processes. With the growing accuracy of measurements, this model becomes not fully adequate (cannot be separated out from the process of measurement), and a unique determination of transition frequency becomes impossible. In the present paper we give an alternative interpretation of the results of the experiment \cite{1}. Though the transition frequency $2s_{1/2}^{F=0}-4p_{1/2}^{F=1}$ cannot be defined with desired accuracy (3 kHz) the main goal of the experiment \cite{1} remains achievable.

The nonuniqueness of determination of transition frequency does not mean that it is impossible to extract the values for various constants (the proton radius and Rydberg constant in case of hydrogen atom) from atomic measurements. For example, the values of the constants can be extracted from the comparison of theoretical and experimental results for any of 3 values given in Eq. (\ref{three}). For accuracy of measurement of the line profile for a certain transition (or its characteristics such as position of the maximum) there are no limits apart from the quantum mechanical uncertainty relations. These measurements should be compared with the theoretical calculations of line shape where the NR corrections are taken into account.


\appendix
\renewcommand{\theequation}{A\arabic{equation}}
\setcounter{equation}{0}
\section*{Appendix}

In this Appendix we present a most general expression for the photon scattering cross section on the hydrogen atom. In this expression not only fine but also hyperfine structure of the atomic levels is taken explicitly into account. We also present the cross sections averaged over the incident photon polarizations and summed over the outgoing photon polarizations. The "closed expression" means that the cross section is written in terms of $ 6j $-symbols and simple radial integrals. The angular evaluations are based on the book \cite{11}. The earlier expressions (not in the fully closed form) for this cross section can be found in \cite{9, amaro1, amaro2}. Using these expressions the nonresonant corrections to the various photon scattering experiments are derived. 

We start from the investigation of angular dependence of the photon scattering cross section on an atom in the experiment of the first type, i.e. when summation over photon polarizations is performed for the differential cross section $ \sum\limits_{\vec{e}_i\vec{e}_f}\sigma_{if} $. Our goal is to derive an analytical expression for the differential cross section with explicit dependence on the angle between directions of absorbed and emitted photons. Before proceeding to the evaluation of these expressions we give the basic relations that we will use below following the notations in \cite{11}.

Summation over photon polarizations can be performed with the use of the formula \cite{10}.
\begin{eqnarray}
\label{r2}
\sum_{\vec{e}}(\vec{e}^{\;*}\vec{a})(\vec{e}\,\vec{b})=(\vec{\nu}\times \vec{a})(\vec{\nu}\times \vec{b}),
\end{eqnarray}
where $ \vec{a} $, $ \vec{b} $ are two arbitrary vectors. We denote the vector components in a cyclic basis as $ (\vec{a})_{q} $, $ q=0,\;\pm 1 $. In general, we will use irreducible tensors $ a_{p} $ of the rank $ p $ with the components $ a_{pq} $. The first lower index denotes the rank and the second one denotes the component. The irreducible tensor $ a_{1} $ of the rank 1 with the components $ a_{1q} $ correspond to the vector $ \vec{a} $ and cyclic vector component $ (\vec{a})_{q} $. The vector component $ (\vec{a})_{q} $  equals the tensor component $ a_{1q} $. Then the vector product of two vectors $ \vec{a} $ and $ \vec{b} $ can be expressed as
\begin{eqnarray}
\label{r1}
\left(\vec{a}\times \vec{b} \right)_{q}=-\mathrm{i}\sqrt{2}\lbrace a_{1}\otimes b_{1}\rbrace_{1q} 
\\\nonumber
=-\mathrm{i}\sqrt{2}\sum_{\mu\nu}C^{1q}_{1\mu 1\nu}a_{\mu}b_{\nu}
\\\nonumber
=-\mathrm{i}\sqrt{6}(-1)^q\sum_{\mu\nu}
\begin{pmatrix}
1 & 1 & 1\\
\mu & \nu & -q
\end{pmatrix}
a_{\mu}b_{\nu}
.
\end{eqnarray}
Here $\otimes$ denotes the tensor product, $ \mu,\;\nu=0,\;\pm 1 $ and $ C^{1q}_{1\mu 1\nu} $ is the Clebsch-Gordan coefficient. Complex conjugation for tensor components:
\begin{eqnarray}
\label{r3}
(a_{yz})^*=(-1)^{-z}a_{y-z}=(-1)^{z}a_{y-z},
\end{eqnarray}
A scalar product of two irreducible tensors of rank $ y $:
\begin{eqnarray}
\label{r4}
\sum_{z}a_{yz}b^*_{yz}=(a_y\cdot b_{y})
\\\nonumber
=(-1)^{-y}\sqrt{2y+1}\left\lbrace a_y\otimes b_{y}\right\rbrace_{00}
.
\end{eqnarray}
Components of tensor product of rank $ y $ of two first rank tensors $ a_{1} $ and $ b_{1} $
\begin{eqnarray}
\label{r5}
\lbrace a_{1}\otimes b_{1}\rbrace_{yz}=(-1)^{z}\sqrt{2y+1}
\\\nonumber
\times\sum_{\mu\nu}
\begin{pmatrix}
1 & 1 & y \\
\mu & \nu & -z
\end{pmatrix}
a_{\mu}b_{\nu}
,
\end{eqnarray}
\begin{eqnarray}
\label{r6}
(\lbrace a_{1}\otimes b_{1}\rbrace_{yz})^{*}=(-1)^{-z}\lbrace a_{1}\otimes b_{1}\rbrace_{y-z}
\\\nonumber
=\sqrt{2y+1}\sum_{\mu\nu}
\begin{pmatrix}
1 & 1 & y \\
\mu & \nu & z
\end{pmatrix}
a_{\mu}b_{\nu}
.
\end{eqnarray}

Now we perform angular algebra in matrix elements for the experiment of the first type, i.e., we calculate $ \sum\limits_{\vec{e}_i\vec{e}_f}\sigma_{if}$. Taking the square modulus of matrix elements in the numerator of Eq. (\ref{3}) of the main text we find
\begin{eqnarray}
\sum\limits_{\vec{e}_i\vec{e}_f}\sigma_{if}=\sum\limits_{\vec{e}_i\vec{e}_f}\sigma_{if}^{\rm res}+\sum_{\vec{e}_i\vec{e}_f}\sigma_{if}^{\rm nr}
,
\end{eqnarray}
where $\sum\limits_{\vec{e}_i\vec{e}_f}\sigma_{if}^{\rm res}$ is the resonant contribution to the differential cross section 
\begin{widetext}
\begin{eqnarray}
\label{r7}
\sum_{\vec{e}_i\vec{e}_f}\sigma_{if}^{\rm res}=\frac{1}{2F_{i}+1}
\sum_{\vec{e}_i\vec{e}_f}\sum_{\substack{M_{F_i}M_{F_f}}}\sum\limits_{nljFM_{F}}
(E_{nljF}-E_{n_i l_i j_i F_i})^{3}(E_{nljF}-E_{n_f l_f j_f F_f})^{3}
\\\nonumber
\times 
\left|
\frac{
\langle n_i l_i j_i F_i M_{F_i} |\vec{e}_{i}\vec{d}|n l j F M_{F}\rangle 
\langle n l j F M_{F}|\vec{e}_{f}^{\;*}\vec{d}|n_f l_f j_f F_f M_{F_f} \rangle
}{E_{nljF}-E_{n_il_ij_iF_i}-\omega-\frac{\mathrm{i}}{2}\Gamma_{nljF}}
\right|^2 d\omega 
,
\end{eqnarray}
and $\sum\limits_{\vec{e}_i\vec{e}_f}\sigma_{if}^{\rm nr}$ is the nonresonant contribution to the cross section
\begin{eqnarray}
\label{r8}
\sum_{\vec{e}_i\vec{e}_f}\sigma_{if}^{\rm nr}=
\frac{1}{2F_{i}+1}
\sum_{\vec{e}_i\vec{e}_f}
\sum_{M_{F_i}M_{F_f}}
\sum_{\substack{n l j  F M_{F} \\ n' l' j' F' M_{F'}}}
(E_{nljF}-E_{n_i l_i j_i F_i})^{3/2}
(E_{n'l'j'F'}-E_{n_i l_i j_i F_i})^{3/2}
\\\nonumber
\times
(E_{nljF}-E_{n_f l_f j_f F_f})^{3/2}
(E_{n'l'j'F'}-E_{n_f l_f j_f F_f})^{3/2}
\\\nonumber
\times
2\mathrm{Re}
\left[
\frac{\langle n_i l_i j_i F_i M_{F_i} |\vec{e}_{i}\vec{d}|n l j F M_{F}\rangle 
\langle n l j F M_{F}|\vec{e}_{f}^{\;*}\vec{d}|n_f l_f j_f F_f M_{F_f} \rangle
}{(E_{nljF}-E_{n_il_ij_iF_i}-\omega-\frac{\mathrm{i}}{2}\Gamma_{nljF})}
\right.
\\\nonumber
\left.
\times
\frac{\langle n' l' j' F' M_{F'}|\vec{e}_{i}^{\;*}\vec{d}|n_i l_i j_i F_i M_{F_i} \rangle 
\langle n_f l_f j_f F_f M_{F_f}  |\vec{e}_{f}\vec{d}|n' l' j' F' M_{F'}\rangle}{(E_{n'l'j'F'}-E_{nljF})}
\right]
d\omega
.
\end{eqnarray}
Performing summation over polarizations in Eq. (\ref{r8}) with the use of Eq. (\ref{r2}) yields
\begin{eqnarray}
\label{r9}
\sum_{\vec{e}_i\vec{e}_f}\sigma_{if}^{\rm nr}
=
\frac{1}{2F_{i}+1}
\sum_{M_{F_i}M_{F_f}}
\sum_{\substack{n l j  F M_{F} \\ n' l' j' F' M_{F'}}}
(E_{nljF}-E_{n_i l_i j_i F_i})^{3/2}
(E_{n'l'j'F'}-E_{n_i l_i j_i F_i})^{3/2}
\\\nonumber
\times
(E_{nljF}-E_{n_f l_f j_f F_f})^{3/2}
(E_{n'l'j'F'}-E_{n_f l_f j_f F_f})^{3/2}
\\\nonumber
\times
2\mathrm{Re}
\left[
\frac{
\langle n_i l_i j_i F_i M_{F_i} |\vec{\nu}_{i}\times\vec{d}|n l j F M_{F}\rangle 
\langle n' l' j' F' M_{F'}|\vec{\nu}_{i}\times\vec{d}|n_i l_i j_i F_i M_{F_i} \rangle 
}{(E_{nljF}-E_{n_il_ij_iF_i}-\omega-\frac{\mathrm{i}}{2}\Gamma_{nljF})}
\right.
\\\nonumber
\times
\left.
\frac{
\langle n_f l_f j_f F_f M_{F_f} |\vec{\nu}_{f}\times\vec{d}|n' l' j' F' M_{F'}\rangle 
\langle n l j F M_{F} |\vec{\nu}_{f}\times\vec{d}|n_f l_f j_f F_f M_{F_f} \rangle 
}{(E_{n'l'j'F'}-E_{nljF})}
\right]
d\omega
.
\end{eqnarray}
The scalar product of two vectors in Eq. (\ref{r9}) can be written in terms of cyclic coordinates 
\begin{eqnarray}
\label{radd9}
\sum_{\vec{e}_i\vec{e}_f}\sigma_{if}^{\rm nr}=
\frac{1}{2F_{i}+1}
\sum_{M_{F_i}M_{F_f}}
\sum_{\substack{n l j  F M_{F}\\ n' l' j' F' M_{F'}}}
(E_{nljF}-E_{n_i l_i j_i F_i})^{3/2}
(E_{n'l'j'F'}-E_{n_i l_i j_i F_i})^{3/2}
\\\nonumber
\times
(E_{nljF}-E_{n_f l_f j_f F_f})^{3/2}
(E_{n'l'j'F'}-E_{n_f l_f j_f F_f})^{3/2}
\\\nonumber
\times
2\mathrm{Re}
\left[
\sum_{qq'}(-1)^{q+q'}
\frac{
\langle n_i l_i j_i F_i M_{F_i} |(\vec{\nu}_{i}\times\vec{d})_{q}|n l j F M_{F}\rangle 
\langle n' l' j' F' M_{F'}|(\vec{\nu}_{i}\times\vec{d})_{-q}|n_i l_i j_i F_i M_{F_i} \rangle 
}{(E_{nljF}-E_{n_il_ij_iF_i}-\omega-\frac{\mathrm{i}}{2}\Gamma_{nljF})}
\right.
\\\nonumber
\left.
\times
\frac{
\langle n_f l_f j_f F_f M_{F_f} |(\vec{\nu}_{f}\times\vec{d})_{q'}|n' l' j' F' M_{F'}\rangle 
\langle n l j F M_{F} |(\vec{\nu}_{f}\times\vec{d})_{-q'}|n_f l_f j_f F_f M_{F_f} \rangle 
}{(E_{n'l'j'F'}-E_{nljF})}
\right]
d\omega
.
\end{eqnarray}
Using Eq. (\ref{r1}) and applying Eckart-Wigner theorem for the dipole matrix elements Eq. (\ref{radd9}) is reduced to
\begin{eqnarray}
\label{r10}
\sum_{\vec{e}_i\vec{e}_f}\sigma_{if}^{\rm nr}=
\frac{36}{2F_{i}+1}
\sum_{M_{F_i}M_{F_f}}
\sum_{\substack{n l j  F M_{F}\\ n' l' j' F' M_{F'}}}
(E_{nljF}-E_{n_i l_i j_i F_i})^{3/2}
(E_{n'l'j'F'}-E_{n_i l_i j_i F_i})^{3/2}
\\\nonumber
\times
(E_{nljF}-E_{n_f l_f j_f F_f})^{3/2}
(E_{n'l'j'F'}-E_{n_f l_f j_f F_f})^{3/2}
\\\nonumber
\times
\sum\limits_{\substack{\mu'''\mu''\mu'\mu\\\nu'''\nu''\nu'\nu}}
\sum\limits_{q'q}
(-1)^{q+q'}
\begin{pmatrix}
1 & 1 & 1\\
\mu & \nu & -q
\end{pmatrix}
\begin{pmatrix}
1 & 1 & 1\\
\mu' & \nu' & q
\end{pmatrix}
\begin{pmatrix}
1 & 1 & 1\\
\mu'' & \nu'' & -q'
\end{pmatrix}
\\\nonumber
\times
\begin{pmatrix}
1 & 1 & 1\\
\mu''' & \nu''' & q'
\end{pmatrix}
\nu^{i}_{\mu}\nu^{i}_{\mu'}\nu^{f}_{\mu''}\nu^{f}_{\mu'''}
(-1)^{F_i-M_{F_i}+F'-M_{F'}+F_f-M_{F_f}+F-M_{F}}
\\\nonumber
\times
\begin{pmatrix}
F_{i} & 1 & F\\
-M_{F_i} & \nu & M_{F}
\end{pmatrix}
\begin{pmatrix}
F' & 1 & F_{i}\\
-M_{F'}& \nu' & M_{F_i}
\end{pmatrix}
\begin{pmatrix}
F_{f} & 1 & F'\\
-M_{F_f} & \nu'' & M_{F'}
\end{pmatrix}
\begin{pmatrix}
F & 1 & F_{f}\\
-M_{F}& \nu''' & M_{F_f}
\end{pmatrix}
\\\nonumber
\times
2\mathrm{Re}\left[
\frac{
\langle n_i l_i j_i F_i ||d_{1}||n l j F \rangle 
\langle n' l' j' F' ||d_{1}||n_i l_i j_i F_i\rangle 
\langle n_f l_f j_f F_f||d_{1}||n' l' j' F'\rangle 
\langle n l j F ||d_{1}||n_f l_f j_f F_f \rangle 
}
{(E_{nljF}-E_{n_il_ij_iF_i}-\omega-\frac{\mathrm{i}}{2}\Gamma_{nljF})(E_{n'l'j'F'}-E_{nljF})}
\right]
d\omega
.
\end{eqnarray}
The reduced matrix elements in Eq. (\ref{r10}) do not depend on projections of any angular momentum and are given by \cite{11}
\begin{eqnarray}
\label{r11}
\langle n'l'j'F'||d_1||nljF\rangle= (-1)^{j'+j+I+l'+1/2+F}
\Pi_{F'}\Pi_{F}\Pi_{j'}\Pi_{j}
\begin{Bmatrix}
j' & F' & I \\
F  & j  & 1
\end{Bmatrix}
\begin{Bmatrix}
l' & j' & 1/2 \\
j  & l  & 1
\end{Bmatrix}
\langle n' l' || d_{1} || nl \rangle
,
\end{eqnarray}
\end{widetext}
where $ I $ is the nuclear spin ($ I=1/2 $ for hydrogen atom) and
\begin{eqnarray}
\label{r12}
\langle n' l' || d_{1} || nl \rangle = e(-1)^{l'}\Pi_{l}\Pi_{l'}
\begin{pmatrix}
l & 1 & l'\\
0 & 0 & 0
\end{pmatrix}
\\\nonumber
\times
\int_{0}^{\infty}r^3 R_{n'l'}R_{nl}dr.
\end{eqnarray}
Here $ R_{nl} $ is the radial part of hydrogen wave function and $ \Pi_{a}=\sqrt{2a+1} $.

Then summations over projections of final and intermediate states and averaging over projections of initial state is performed independently with the use of Eq. (10) in section 12.1 of \cite{11}
\begin{widetext}
\begin{eqnarray}
\label{r13}
\sum_{M_{F_i}M_{F}M_{F'}M_{F_f}}
(-1)^{F_i-M_{F_i}+F'-M_{F'}+F_f-M_{F_f}+F-M_{F}}
\begin{pmatrix}
F_{i} & 1 & F\\
-M_{F_i} & \nu & M_{F}
\end{pmatrix}
\begin{pmatrix}
F' & 1 & F_{i}\\
-M_{F'}& \nu' & M_{F_i}
\end{pmatrix}
\begin{pmatrix}
F_{f} & 1 & F'\\
-M_{F_f} & \nu'' & M_{F'}
\end{pmatrix}
\\\nonumber
\times
\begin{pmatrix}
F & 1 & F_{f}\\
-M_{F}& \nu''' & M_{F_f}
\end{pmatrix}
=(-1)^{F'-F}
\sum_{x\xi}(-1)^{x-\xi}\Pi_{x}^2
\begin{pmatrix}
1 & x & 1 \\
-\nu & -\xi & -\nu'
\end{pmatrix}
\begin{pmatrix}
1 & x & 1 \\
-\nu''' & \xi & -\nu''
\end{pmatrix}
\\\nonumber
\times
\begin{Bmatrix}
1 & x & 1 \\
F' & F_{i} & F
\end{Bmatrix}
\begin{Bmatrix}
1 & x & 1 \\
F' & F_{f} & F
\end{Bmatrix}
.
\end{eqnarray}
\end{widetext}
Now we can consider the sum over indices $ \nu''\nu'''q' $. This summation is performed independently on indices $ \nu\nu'q $ with the use of Eq. (6) in section 12.1 of \cite{11}. Terms in Eq. (\ref{r10}) depending on variables $ \nu''\nu'''q' $ are
\begin{widetext}
\begin{eqnarray}
\label{r14}
\sum_{\nu''\nu'''}\sum_{q'}
(-1)^{q'}
(-1)^{x-\xi}
\begin{pmatrix}
1 & 1 & 1\\
\mu'' & \nu'' & -q'
\end{pmatrix}
\begin{pmatrix}
1 & 1 & 1\\
\mu''' & \nu''' & q'
\end{pmatrix}
\begin{pmatrix}
1 & x & 1 \\
-\nu''' & \xi & -\nu''
\end{pmatrix}
\\\nonumber
=
\sum_{\nu''\nu'''}\sum_{q'}
(-1)^{1-q'+1-\nu'' +1-\nu'''}
\begin{pmatrix}
1 & 1 & 1\\
q' & \mu''' & -\nu'''
\end{pmatrix}
\begin{pmatrix}
1 & x & 1 \\
\nu''' & \xi & -\nu''
\end{pmatrix}
\begin{pmatrix}
1 & 1 & 1\\
\nu'' & \mu'' & -q'
\end{pmatrix}
\\\nonumber
=
(-1)^{x}
\begin{pmatrix}
1 & x & 1 \\
-\mu''' & -\xi & -\mu''
\end{pmatrix}
\begin{Bmatrix}
1 & x & 1\\
1 & 1 & 1
\end{Bmatrix}
.
\end{eqnarray}
Another sum over indices $ \nu\nu'q $ is reduced in a similar way as follows 
\begin{eqnarray}
\label{r15}
\sum_{\nu\nu'}\sum_{q}
(-1)^{q}
\begin{pmatrix}
1 & 1 & 1\\
\mu & \nu & -q
\end{pmatrix}
\begin{pmatrix}
1 & 1 & 1\\
\mu' & \nu' & q
\end{pmatrix}
\begin{pmatrix}
1 & x & 1 \\
-\nu & -\xi & -\nu'
\end{pmatrix}
\\\nonumber
=
(-1)^{x-\xi}
\sum_{\nu\nu'}\sum_{q}
(-1)^{1-q+1-\nu'+1-\nu}
\begin{pmatrix}
1 & 1 & 1\\
 q & \mu' & -\nu' 
\end{pmatrix}
\begin{pmatrix}
1 & x & 1 \\
\nu' & -\xi & -\nu
\end{pmatrix}
\begin{pmatrix}
1 & 1 & 1\\
\nu & \mu & -q
\end{pmatrix}
\\\nonumber
=(-1)^{x-\xi}
\begin{pmatrix}
1 & x & 1 \\
-\mu' & \xi & -\mu
\end{pmatrix}
\begin{Bmatrix}
1 & x & 1\\
1 & 1 & 1
\end{Bmatrix}
.
\end{eqnarray}
Collecting together results of Eqs. (\ref{r13}), (\ref{r14}), (\ref{r15}) yields
\begin{eqnarray}
\label{r16}
\sum_{\vec{e}_i\vec{e}_f}\sigma_{if}^{\rm nr}=
\sum_{\substack{n l j  F \\ n' l' j' F' }}
\frac{36(-1)^{F'-F}}{2F_{i}+1}
(E_{nljF}-E_{n_i l_i j_i F_i})^{3/2}
(E_{n'l'j'F'}-E_{n_i l_i j_i F_i})^{3/2}
\\\nonumber
\times
(E_{nljF}-E_{n_f l_f j_f F_f})^{3/2}
(E_{n'l'j'F'}-E_{n_f l_f j_f F_f})^{3/2}
\\\nonumber
\times
\sum_{x}\sum_{\xi}\sum_{\mu\mu'\mu''\mu'''}
(-1)^{2x-\xi}
\Pi_{x}^2
\begin{pmatrix}
1 & x & 1 \\
-\mu' & \xi & -\mu
\end{pmatrix}
\begin{pmatrix}
1 & x & 1 \\
-\mu''' & -\xi & -\mu''
\end{pmatrix}
\begin{Bmatrix}
1 & x & 1\\
1 & 1 & 1
\end{Bmatrix}
^2
\\\nonumber
\times
\nu^{i}_{\mu}\nu^{i}_{\mu'}\nu^{f}_{\mu''}\nu^{f}_{\mu'''}
\begin{Bmatrix}
1 & x & 1 \\
F' & F_{i} & F
\end{Bmatrix}
\begin{Bmatrix}
1 & x & 1 \\
F' & F_{f} & F
\end{Bmatrix}
\sum_{\substack{nljF\\n'l'j'F'}}
\\\nonumber
\times
2\mathrm{Re}
\left[
\frac{
\langle n_i l_i j_i F_i ||d_{1}||n l j F \rangle 
\langle n l j F ||d_{1}||n_f l_f j_f F_f\rangle 
\langle n_i l_i j_i F_i||d_{1}||n' l' j' F'\rangle 
\langle n' l' j' F' ||d_{1}||n_f l_f j_f F_f \rangle 
}
{(E_{nljF}-E_{n_il_ij_iF_i}-\omega-\frac{\mathrm{i}}{2}\Gamma_{nljF})(E_{n'l'j'F'}-E_{nljF})}
\right]
d\omega 
.
\end{eqnarray}
\end{widetext}
Two remaining $ 3jm $ symbols in Eq. (\ref{r16}) can be considered separately. Using Eq. (5) in section 12.1 of \cite{11} and Eqs. (\ref{r3})-(\ref{r6}), the following sequence of equalities can be written for the sum over $ \xi $ in Eq. (\ref{r16}) 
\begin{widetext}
\begin{eqnarray}
\label{r17}
\sum_{\xi}
(-1)^{-\xi}
\begin{pmatrix}
1 & x & 1 \\
-\mu' & \xi & -\mu
\end{pmatrix}
\begin{pmatrix}
1 & x & 1 \\
-\mu''' & -\xi & -\mu''
\end{pmatrix}
\nu^{i}_{\mu}\nu^{i}_{\mu'}\nu^{f}_{\mu''}\nu^{f}_{\mu'''}
\\\nonumber
=\sum_{yz}(-1)^{x}
\left(\left\lbrace \nu^{i}_{1}\otimes\nu^{f}_{1}  \right\rbrace_{yz}\right)^{*} \left\lbrace \nu^{i}_{1}\otimes\nu^{f}_{1}  \right\rbrace_{yz}
\begin{Bmatrix}
1 & 1 & y \\
1 & 1 & x
\end{Bmatrix}
\\\nonumber
=\sum_{y}(-1)^{x-y}\sqrt{2y+1}
\left\lbrace
\left\lbrace \nu^{i}_{1}\otimes\nu^{f}_{1}  \right\rbrace_{yz}\otimes \left\lbrace \nu^{i}_{1}\otimes\nu^{f}_{1}  \right\rbrace_{yz}
\right\rbrace_{00}
\begin{Bmatrix}
1 & 1 & y \\
1 & 1 & x
\end{Bmatrix}
\end{eqnarray}
Substitution of Eq. (\ref{r17}) into Eq. (\ref{r16}) yields
\begin{eqnarray}
\label{r18}
\sum_{\vec{e}_i\vec{e}_f}\sigma_{if}^{\rm res}=
\frac{f^{(1)}_{\rm res}}
{(E_{nljF}-E_{n_il_ij_iF_i}-\omega)^2-\frac{\Gamma_{nljF}^2}{4}}
d\omega 
,
\end{eqnarray}

\begin{eqnarray}
\label{r19}
\sum_{\vec{e}_i\vec{e}_f}\sigma_{if}^{\rm nr}=
2\mathrm{Re}\left[
\frac{f^{(1)}_{\rm nr}}
{(E_{nljF}-E_{n_il_ij_iF_i}-\omega-\frac{\mathrm{i}}{2}\Gamma_{nljF})(E_{n'l'j'F'}-E_{nljF})}
\right]
d\omega 
,
\end{eqnarray}
together with notations
\begin{eqnarray}
\label{r20}
f^{(1)}_{\rm res}=
(E_{nljF}-E_{n_i l_i j_i F_i})^{3}(E_{nljF}-E_{n_f l_f j_f F_f})^{3}
\sum\limits_{xy}A_{xy}^{(1)\;\mathrm{res}}
\left\lbrace
\left\lbrace \nu^{i}_{1}\otimes\nu^{f}_{1}  \right\rbrace_{y}\otimes \left\lbrace \nu^{i}_{1}\otimes\nu^{f}_{1}  \right\rbrace_{y}
\right\rbrace_{00}
\\\nonumber
,
\end{eqnarray}
\begin{eqnarray}
\label{r21}
f^{(1)}_{\rm nr}=
(E_{nljF}-E_{n_i l_i j_i F_i})^{3/2}
(E_{n'l'j'F'}-E_{n_i l_i j_i F_i})^{3/2}
(E_{nljF}-E_{n_f l_f j_f F_f})^{3/2}
(E_{n'l'j'F'}-E_{n_f l_f j_f F_f})^{3/2}
\\\nonumber
\times
\sum\limits_{xy}A_{xy}^{(1)\;\mathrm{nr}}
\left\lbrace
\left\lbrace \nu^{i}_{1}\otimes\nu^{f}_{1}  \right\rbrace_{y}\otimes \left\lbrace \nu^{i}_{1}\otimes\nu^{f}_{1}  \right\rbrace_{y}
\right\rbrace_{00}
\end{eqnarray}
\begin{eqnarray}
A_{xy}^{(1)\;\mathrm{res}}=
\frac{36(-1)^{x-y}}{2F_{i}+1}
\Pi_{x}^2\Pi_{y}
\begin{Bmatrix}
1 & 1 & y \\
1 & 1 & x
\end{Bmatrix}
\begin{Bmatrix}
1 & x & 1 \\
1 & 1 & 1
\end{Bmatrix}
^2
\begin{Bmatrix}
1 & x & 1 \\
F & F_{i} & F
\end{Bmatrix}
\begin{Bmatrix}
1 & x & 1 \\
F & F_{f} & F
\end{Bmatrix}
\\\nonumber
\times
|\langle n_i l_i j_i F_i ||d_{1}||n l j F \rangle 
\langle n l j F ||d_{1}||n_f l_f j_f F_f\rangle |^2
\end{eqnarray}

\begin{eqnarray}
A_{xy}^{(1)\;\mathrm{nr}}=
\frac{36(-1)^{F'-F+x-y}}{2F_{i}+1}
\Pi_{x}^2\Pi_{y}
\begin{Bmatrix}
1 & 1 & y \\
1 & 1 & x
\end{Bmatrix}
\begin{Bmatrix}
1 & x & 1 \\
1 & 1 & 1
\end{Bmatrix}
^2
\begin{Bmatrix}
1 & x & 1 \\
F' & F_{i} & F
\end{Bmatrix}
\begin{Bmatrix}
1 & x & 1 \\
F' & F_{f} & F
\end{Bmatrix}
\\\nonumber
\times
\langle n_i l_i j_i F_i ||d_{1}||n l j F \rangle 
\langle n' l' j' F' ||d_{1}||n_i l_i j_i F_i\rangle 
\langle n_f l_f j_f F_f||d_{1}||n' l' j' F'\rangle 
\langle n l j F ||d_{1}||n_f l_f j_f F_f \rangle 
.
\end{eqnarray}

Now we can consider the second type of experiment, i.e. differential cross section $ \sum\limits_{\vec{e}_f}\sigma_{if}^{\rm nr} $. Taking square of modulus in Eq. (\ref{r9}), performing summation over photon polarization vector $ \vec{e}_f $, summation over projections in final states and averaging over projections in 
\begin{eqnarray}
\label{r22}
\sum_{\vec{e}_i}\sigma_{if}^{\rm nr}=\frac{1}{2F_{i}+1}
\sum\limits_{M_{F_f}M_{F_i}}
\sum_{\substack{n l j  F M_{F}\\n' l' j' F' M_{F'}}}
(E_{nljF}-E_{n_i l_i j_i F_i})^{3/2}
(E_{n'l'j'F'}-E_{n_i l_i j_i F_i})^{3/2}
\\\nonumber
\times
(E_{nljF}-E_{n_f l_f j_f F_f})^{3/2}
(E_{n'l'j'F'}-E_{n_f l_f j_f F_f})^{3/2}
\\\nonumber
\times
2\mathrm{Re}
\left[
\frac{
\langle n_i l_i j_i F_i M_{F_i} |\vec{e}_{i}\vec{d}|n l j F M_{F}\rangle 
\langle n' l' j' F' M_{F'}|\vec{e}_{i}^{\;*}\vec{d}|n_i l_i j_i F_i M_{F_i} \rangle 
}{(E_{nljF}-E_{n_il_ij_iF_i}-\omega-\frac{\mathrm{i}}{2}\Gamma_{nljF})}
\right.
\\\nonumber
\left.
\times
\frac{
\langle n_f l_f j_f F_f M_{F_f}  |\vec{\nu}_{f}\times \vec{d}|n' l' j' F' M_{F'}\rangle
\langle n l j F M_{F}|\vec{\nu}_{f}\times \vec{d}|n_f l_f j_f F_f M_{F_f} \rangle
}
{(E_{n'l'j'F'}-E_{nljF})}
\right]
d\omega 
.
\end{eqnarray}
The scalar product of two vectors in Eq. (\ref{r22}) can be written in cyclic coordinate space as
\begin{eqnarray}
\label{r23}
\sum_{\vec{e}_i}\sigma_{if}^{\rm nr}=
\frac{1}{2F_{i}+1}
\sum\limits_{M_{F_f}M_{F_i}}
\sum_{\substack{n l j  F M_{F}\\n' l' j' F' M_{F'}}}
(E_{nljF}-E_{n_i l_i j_i F_i})^{3/2}
(E_{n'l'j'F'}-E_{n_i l_i j_i F_i})^{3/2}
\\\nonumber
\times
(E_{nljF}-E_{n_f l_f j_f F_f})^{3/2}
(E_{n'l'j'F'}-E_{n_f l_f j_f F_f})^{3/2}
\\\nonumber
\times
2\mathrm{Re}
\sum_{qq'q''} (-1)^{q+q'+q''}
\left[
\frac{
\langle n_i l_i j_i F_i M_{F_i} |e^{i}_qd_{-q}|n l j F M_{F}\rangle 
\langle n' l' j' F' M_{F'}|e^{i,*}_{q'}d_{-q'}|n_i l_i j_i F_i M_{F_i} \rangle 
}{(E_{nljF}-E_{n_il_ij_iF_i}-\omega-\frac{\mathrm{i}}{2}\Gamma_{nljF})}
\right.
\\\nonumber
\times
\left.
\frac{
\langle n_f l_f j_f F_f M_{F_f}  |(\vec{\nu}_{f}\times \vec{d})_{q''}|n' l' j' F' M_{F'}\rangle
\langle n l j F M_{F}|(\vec{\nu}_{f}\times \vec{d})_{-q''}|n_f l_f j_f F_f M_{F_f} \rangle
}{(E_{n'l'j'F'}-E_{nljF})}
\right]
d\omega 
.
\end{eqnarray}
Using Eq. (\ref{r1}) for cyclic components of vector product and applying the Eckart-Wigner theorem for the dipole matrix elements in Eq. (\ref{r23}) we find
\begin{eqnarray}
\label{r24}
\sum_{\vec{e}_i}\sigma_{if}^{\rm nr}=
-\frac{6}{2F_{i}+1}
\sum\limits_{M_{F_f}M_{F_i}}
\sum_{\substack{n l j  F M_{F}\\n' l' j' F' M_{F'}}}
(E_{nljF}-E_{n_i l_i j_i F_i})^{3/2}
(E_{n'l'j'F'}-E_{n_i l_i j_i F_i})^{3/2}
\\\nonumber
\times
(E_{nljF}-E_{n_f l_f j_f F_f})^{3/2}
(E_{n'l'j'F'}-E_{n_f l_f j_f F_f})^{3/2}
\sum\limits_{\substack{\mu'\mu\\ \nu'\nu}}
\sum\limits_{q''q'q}
(-1)^{q+q'+q''}(-1)^{F_i-M_{F_i}+F'-M_{F'}+F_f-M_{F_f}+F-M_{F}}
\\\nonumber
\times
\begin{pmatrix}
F_{i} & 1 & F\\
-M_{F_i} & -q & M_{F}
\end{pmatrix}
\begin{pmatrix}
F' & 1 & F_{i}\\
-M_{F'}& -q' & M_{F_i}
\end{pmatrix}
\begin{pmatrix}
F_{f} & 1 & F'\\
-M_{F_f} & \nu & M_{F'}
\end{pmatrix}
\\\nonumber
\times
\begin{pmatrix}
F & 1 & F_{i}\\
-M_{F}& \nu' & M_{F_i}
\end{pmatrix}
\begin{pmatrix}
1 & 1 & 1\\
\mu & \nu & -q''
\end{pmatrix}
\begin{pmatrix}
1 & 1 & 1\\
\mu' & \nu' & q''
\end{pmatrix}
e^{i}_{q}e^{i,*}_{q'}\nu^{f}_{\mu}\nu^{f}_{\mu'}
\\\nonumber
\times
2\mathrm{Re}
\left[
\frac{
\langle n_i l_i j_i F_i ||d_{1}||n l j F \rangle 
\langle n' l' j' F' ||d_{1}||n_i l_i j_i F_i\rangle 
\langle n_f l_f j_f F_f||d_{1}||n' l' j' F'\rangle 
\langle n l j F ||d_{1}||n_f l_f j_f F_f \rangle 
}{(E_{nljF}-E_{n_il_ij_iF_i}-\omega-\frac{\mathrm{i}}{2}\Gamma_{nljF})(E_{n'l'j'F'}-E_{nljF})}
\right]
d\omega 
.
\end{eqnarray}
Then summation over projections of the angular momentum in the final, intermediate and initial states in Eq. (\ref{r24}) is performed independently on the reduced matrix elements and vector components:
\begin{eqnarray}
\label{r25}
\sum_{M_{F_i}M_{F'}M_{F}M_{F_f}}(-1)^{F_i-M_{F_i}+F'-M_{F'}+F_f-M_{F_f}+F-M_{F}}
\\\nonumber
\times
\begin{pmatrix}
F_{i} & 1 & F\\
-M_{F_i} & -q & M_{F}
\end{pmatrix}
\begin{pmatrix}
F' & 1 & F_{i}\\
-M_{F'}& -q' & M_{F_i}
\end{pmatrix}
\begin{pmatrix}
F_{f} & 1 & F'\\
-M_{F_f} & \nu & M_{F'}
\end{pmatrix}
\begin{pmatrix}
F & 1 & F_{f}\\
-M_{F}& \nu' & M_{F_f}
\end{pmatrix}
\\\nonumber
=
(-1)^{F'-F}(-1)^{q+q'+q''}
\sum_{x\xi}(-1)^{x-\xi}\Pi_{x}^2
\begin{pmatrix}
1 & x & 1 \\
q	 & -\xi & q'
\end{pmatrix}
\begin{pmatrix}
1 & x & 1 \\
-\nu' & \xi & -\nu
\end{pmatrix}
\begin{Bmatrix}
1 & x & 1 \\
F' & F_{i} & F
\end{Bmatrix}
\begin{Bmatrix}
1 & x & 1 \\
F' & F_{f} & F
\end{Bmatrix}
\end{eqnarray}
Substitution of Eq. (\ref{r25}) into equation (\ref{r24}) yields
\begin{eqnarray}
\label{r27}
\sum_{\vec{e}_i}\sigma_{if}^{\rm nr}=
\sum_{\substack{nljF\\n'l'j'F'}}
\frac{6(-1)^{F'-F}}{2F_{i}+1}
(E_{nljF}-E_{n_i l_i j_i F_i})^{3/2}
(E_{n'l'j'F'}-E_{n_i l_i j_i F_i})^{3/2}
\\\nonumber
\times
(E_{nljF}-E_{n_f l_f j_f F_f})^{3/2}
(E_{n'l'j'F'}-E_{n_f l_f j_f F_f})^{3/2}
\sum\limits_{\substack{\mu'\mu \\ \nu'\nu}}\sum\limits_{q''q'q}\sum_{x\xi}(-1)^{q+q'+q''}
\\\nonumber
\times
(-1)^{x-\xi}\Pi_{x}^2
\begin{pmatrix}
1 & x & 1 \\
q	 & -\xi & q'
\end{pmatrix}
\begin{pmatrix}
1 & x & 1 \\
-\nu' & \xi & -\nu
\end{pmatrix}
\begin{pmatrix}
1 & 1 & 1\\
\mu & \nu & -q''
\end{pmatrix}
\begin{pmatrix}
1 & 1 & 1\\
\mu' & \nu' & q''
\end{pmatrix}
\begin{Bmatrix}
1 & x & 1 \\
F' & F_{i} & F
\end{Bmatrix}
\begin{Bmatrix}
1 & x & 1 \\
F' & F_{f} & F
\end{Bmatrix}
e^{i}_{q}e^{i,*}_{q'}\nu^{f}_{\mu}\nu^{f}_{\mu'}
\\\nonumber
\times
2\mathrm{Re}\left[
\frac{
\langle n_i l_i j_i F_i ||d_{1}||n l j F \rangle 
\langle n' l' j' F' ||d_{1}||n_i l_i j_i F_i\rangle 
\langle n_f l_f j_f F_f||d_{1}||n' l' j' F'\rangle 
\langle n l j F ||d_{1}||n_f l_f j_f F_f \rangle 
}
{(E_{nljF}-E_{n_il_ij_iF_i}-\omega-\frac{\mathrm{i}}{2}\Gamma_{nljF})(E_{n'l'j'F'}-E_{nljF})}
\right]
d\omega
.
\end{eqnarray}
Summation over indices $ q''\xi\nu\nu' $ in Eq. (\ref{r27}) is performed with the use of Eq. (9) in section 12.1 in \cite{11}
\begin{eqnarray}
\label{r28}
\sum_{q''\xi\nu\nu'}(-1)^{q''-\xi}
\begin{pmatrix}
1 & x & 1 \\
-q	 & -\xi & -q'
\end{pmatrix}
\begin{pmatrix}
1 & x & 1 \\
-\nu' & \xi & -\nu
\end{pmatrix}
\begin{pmatrix}
1 & 1 & 1\\
\mu & \nu & -q''
\end{pmatrix}
\begin{pmatrix}
1 & 1 & 1\\
\mu' & \nu' & q''
\end{pmatrix}
\\\nonumber
=-(-1)^x
\begin{Bmatrix}
1 & 1 & x \\
1 & 1 & 1
\end{Bmatrix}
\sum_{\xi}(-1)^{x-\xi}
\begin{pmatrix}
1 & 1 & x \\
q' & q & -\xi
\end{pmatrix}
\begin{pmatrix}
x & 1 & 1 \\
\xi & \mu & \mu'
\end{pmatrix}
\end{eqnarray}
Then using Eq. (5) in section 12.1 of \cite{11} for the sum over $ \xi $, Eq. (\ref{r28}) reduces to
\begin{eqnarray}
\label{r29}
-(-1)^x
\begin{Bmatrix}
1 & 1 & x \\
1 & 1 & 1
\end{Bmatrix}
\sum_{\xi}(-1)^{x-\xi}
\begin{pmatrix}
1 & 1 & x \\
q' & q & -\xi
\end{pmatrix}
\begin{pmatrix}
x & 1 & 1 \\
\xi & \mu & \mu'
\end{pmatrix}
\\\nonumber
=
-(-1)^x
\begin{Bmatrix}
1 & 1 & x \\
1 & 1 & 1
\end{Bmatrix}
\sum_{yz}(-1)^{y-z}
\sqrt{2y+1}
\begin{pmatrix}
1 & 1 & y \\
q' & \mu' & -z
\end{pmatrix}
\sqrt{2y+1}
\begin{pmatrix}
 1 & 1 & y\\
\mu & q & z
\end{pmatrix}
\begin{Bmatrix}
1 & 1 & y \\
1 & 1 & x
\end{Bmatrix}
.
\end{eqnarray}
Finally, taking into account and Eqs. (\ref{r3})-(\ref{r6}) and substituting result Eq. (\ref{r29}) into Eq. (\ref{r25}) we find
\begin{eqnarray}
\label{r30}
\sum_{\vec{e}_i}\sigma_{if}^{\rm res}=
\frac{f^{(2)}_{\rm res}}
{(E_{nljF}-E_{n_il_ij_iF_i}-\omega)^2-\frac{\Gamma_{nljF}^2}{4}}d\omega 
,
\end{eqnarray}
\begin{eqnarray}
\label{r31}
\sum_{\vec{e}_i}\sigma_{if}^{\rm nr}=2\mathrm{Re}\left[
\frac{f^{(2)}_{\rm nr}}
{{(E_{nljF}-E_{n_il_ij_iF_i}-\omega-\frac{\mathrm{i}}{2}\Gamma_{nljF})(E_{n'l'j'F'}-E_{nljF})}}
\right]d\omega 
,
\end{eqnarray}
where the following notations are introduced
\begin{eqnarray}
\label{r32}
f^{(2)}_{\rm res}=
(E_{nljF}-E_{n_i l_i j_i F_i})^{3}(E_{nljF}-E_{n_f l_f j_f F_f})^{3}
\sum_{xy}A_{xy}^{(2)\;\mathrm{res}}
\left\lbrace
\left\lbrace e^{i}_{1}\otimes\nu^{f}_{1}  \right\rbrace_{y}\otimes \left\lbrace e^{i}_{1}\otimes\nu^{f}_{1}  \right\rbrace_{y}
\right\rbrace_{00}
,
\end{eqnarray}
\begin{eqnarray}
\label{r33}
f^{(2)}_{\rm nr}=(E_{nljF}-E_{n_i l_i j_i F_i})^{3/2}
(E_{n'l'j'F'}-E_{n_i l_i j_i F_i})^{3/2}
(E_{nljF}-E_{n_f l_f j_f F_f})^{3/2}
(E_{n'l'j'F'}-E_{n_f l_f j_f F_f})^{3/2}
\\\nonumber
\times
\sum_{xy}A_{xy}^{(2)\;\mathrm{\rm nr}}
\left\lbrace
\left\lbrace e^{i}_{1}\otimes\nu^{f}_{1}  \right\rbrace_{y}\otimes \left\lbrace e^{i}_{1}\otimes\nu^{f}_{1}  \right\rbrace_{y}
\right\rbrace_{00}
,
\end{eqnarray}
\begin{eqnarray}
A_{xy}^{(2)\;\mathrm{res}}=\frac{6(-1)^{-y}}{2F_{i}+1}\Pi_{x}^2\Pi_{y}
\begin{Bmatrix}
1 & 1 & y \\
1 & 1 & x
\end{Bmatrix}
\begin{Bmatrix}
1 & 1 & x \\
1 & 1 & 1
\end{Bmatrix}
\begin{Bmatrix}
1 & x & 1 \\
F & F_{i} & F
\end{Bmatrix}
\begin{Bmatrix}
1 & x & 1 \\
F & F_{f} & F
\end{Bmatrix}
\\\nonumber
\times
|
\langle n_i l_i j_i F_i ||d_{1}||n l j F \rangle 
\langle n l j F ||d_{1}||n_f l_f j_f F_f \rangle |^2
,
\end{eqnarray}
\begin{eqnarray}
\label{r34}
A_{xy}^{(2)\;\mathrm{nr}}=\frac{6(-1)^{F'-F-y}}{2F_{i}+1}\Pi_{x}^2\Pi_{y}
\begin{Bmatrix}
1 & 1 & y \\
1 & 1 & x
\end{Bmatrix}
\begin{Bmatrix}
1 & 1 & x \\
1 & 1 & 1
\end{Bmatrix}
\begin{Bmatrix}
1 & x & 1 \\
F' & F_{i} & F
\end{Bmatrix}
\begin{Bmatrix}
1 & x & 1 \\
F' & F_{f} & F
\end{Bmatrix}
\\\nonumber
\times
\langle n_i l_i j_i F_i ||d_{1}||n l j F \rangle 
\langle n' l' j' F' ||d_{1}||n_i l_i j_i F_i\rangle 
\langle n_f l_f j_f F_f||d_{1}||n' l' j' F'\rangle 
\langle n l j F ||d_{1}||n_f l_f j_f F_f \rangle 
.
\end{eqnarray}
\end{widetext}
Tensor product in Eqs. (\ref{r20}), (\ref{r21}), (\ref{r32}), (\ref{r33}) can be expressed through trigonometric functions of the angle between the vectors $ \vec{a} $, $ \vec{b} $. The term with $ y=0 $ in the last factor in Eqs. (\ref{r20}), (\ref{r21}), (\ref{r32}), (\ref{r33}) reduces to the square of scalar product of two vectors $ \vec{a} $ and $ \vec{b} $ \cite{11}
\begin{eqnarray}
\label{r35}
\left\lbrace\left\lbrace a_{1}^{(1,2)}\otimes b_{1}^{(1,2)} \right\rbrace_{0}\otimes
\left\lbrace a_{1}^{(1,2)}\otimes b_{1}^{(1,2)} \right\rbrace_{0}\right\rbrace_{00}
\\\nonumber
=\frac{1}{3}\mathrm{cos}^2\theta_{(1,2)}
,
\end{eqnarray}
where $ a^{(1)}_1=\nu^{i}_{1} $, $ a^{(2)}_1=e^{i}_{1} $, $b^{(1)}_1=b^{(2)}_1=\nu^{f}_{1}$ and $ \theta_{(1,2)} $ is the angle between vectors $ \vec{a} $ and $ \vec{b} $. The term with $ y=1 $ in Eqs. (\ref{r20}), (\ref{r21}), (\ref{r32}), (\ref{r33}) reduces to the square of the vector product of vectors $ \vec{a} $ and $ \vec{b} $
\begin{eqnarray}
\label{r36}
\left\lbrace\left\lbrace a_{1}^{(1,2)}\otimes b_{1}^{(1,2)} \right\rbrace_{1}\otimes
\left\lbrace a_{1}^{(1,2)}\otimes b_{1}^{(1,2)} \right\rbrace_{1}\right\rbrace_{00}
\\\nonumber
=\frac{1}{2\sqrt{3}}\mathrm{sin}^2\theta_{(1,2)}
.
\end{eqnarray} 
The term with $ y=2 $ reduces to the scalar product of the two irreducible tensors of the rank $ 2 $
\begin{eqnarray}
\label{r37}
\left\lbrace\left\lbrace a_{1}^{(1,2)}\otimes b_{1}^{(1,2)} \right\rbrace_{2}\otimes
\left\lbrace a_{1}^{(1,2)}\otimes b_{1}^{(1,2)} \right\rbrace_{2}\right\rbrace_{00}
\\\nonumber
=
\frac{1}{6\sqrt{5}}
(3+\mathrm{cos}^2\theta_{(1,2)}).
\end{eqnarray} 
The Eqs. (\ref{r18}), (\ref{r19}), (\ref{r30}), (\ref{r31}) together with Eqs. (\ref{r35})-(\ref{r37}) conclude the derivation of Eqs. (5) and (9) in the main text. 

Now we are in position to find out the dependence on the angle $ \theta_{(1,2)} $ for the NR correction, i.e. for the Eq. (\ref{avr}) in the main text
\begin{eqnarray}
\label{r38}
\delta\omega^{(1,2)} = \frac{\sum\limits_{n_fl_fj_fF_f}f^{(1,2)}_{\rm nr}}{\sum\limits_{n_fl_fj_fF_f}f^{(1,2)}_{\rm res}}\frac{\Gamma^2}{4\Delta}
.
\end{eqnarray}
For this purpose we set in all equations $ n_i l_i=2s $, $ j_i=1/2 $, $ F_i=0 $, $ nl=4p $, $ j=1/2 $, $ F=1 $, $ j'=3/2 $, $ F'=1 $. Then performing summation in Eq. (\ref{r38}) for the experiment of the first type we find
\begin{eqnarray}
\label{r39}
\delta\omega^{(1)} = \frac{5I_{4p1s}^2+5I_{4p2s}^2+5I_{4p3s}^2+I_{4p3d}^2}{40(I_{4p1s}^2+I_{4p2s}^2+I_{4p3s}^2+I_{4p3d}^2)}
\\\nonumber
\times
\frac{\Gamma^2}{4\Delta}
(1+3\mathrm{cos}\;2\theta_{(1)}),
\end{eqnarray}
where 
\begin{eqnarray}
I_{n'l'nl}=\int_{0}^{\infty}r^3 R_{n'l'}R_{nl}dr
,
\end{eqnarray}
and $ R_{nl} $ is the radial part of hydrogen wave function. 
In Eq. (\ref{r39}) we neglected the fine structure dependencies for the energies in Eqs. (\ref{r20}) and (\ref{r21}), as well as the ratio of energies close to 1 for each term of the sum over final states in Eq. (\ref{r38}). In the same way for the second type of the experiment we find
 \begin{eqnarray}
\label{r40}
\delta\omega^{(2)} = -\frac{5I_{4p1s}^2+5I_{4p2s}^2+5I_{4p3s}^2+I_{4p3d}^2}{20(I_{4p1s}^2+I_{4p2s}^2+I_{4p3s}^2+I_{4p3d}^2)}
\\\nonumber
\times
\frac{\Gamma^2}{4\Delta}(1+3\mathrm{cos}\;2\theta_{(2)})
.
\end{eqnarray}
The angular factor in Eqs. (\ref{r39}) and (\ref{r40}) can be also expressed in terms of Legendre polynomial of second order using the equality $ P_{2}(\mathrm{cos}\theta_{(1,2)}) = \frac{1}{4}(1+3\mathrm{cos}\;2\theta_{(1,2)}) $. Solving equation $ 1+3\mathrm{cos}\;2\theta_{(1,2)}=0 $ for the variable $ \theta_{(1,2)} $ one can easily find that the NR correction vanishes at the angles $ \theta_{(1,2)}=\frac{1}{2}(\pm\mathrm{arccos}\frac{1}{3} + 2 \pi n) $ (with an arbitrary integer $ n $). This result can be obtained for any set of initial and intermediate quantum numbers and corresponds to Fig. 1 of the main text.

\end{document}